\documentstyle[aps,twocolumn,epsfig,amssymb]{revtex} 

\def\lsim{\lower.8ex\hbox{$\buildrel<\over\sim$}}

\begin{document} 
\draft
\twocolumn[\hsize\textwidth\columnwidth\hsize\csname
@twocolumnfalse\endcsname
\title{Discrete charge patterns, Coulomb correlations and interactions 
in protein solutions}
\author {E. Allahyarov$^1$, H. L\"{o}wen$^1$, A.A. Louis$^2$, J.P. Hansen$^2$}
 \address{$^1$Institut f\"{u}r Theoretische Physik
II, Heinrich-Heine-Universit\"{a}t D\"{u}sseldorf, D-40225
D\"{u}sseldorf, Germany} 
\address{$^2$Department of Chemistry, Lensfield Rd, Cambridge CB2 1EW,
UK}
 \maketitle
\begin{abstract}
The effective Coulomb interaction between globular proteins is
calculated as a function of monovalent salt concentration $c_s$, by
explicit Molecular Dynamics simulations of pairs of model proteins in
the presence of microscopic co and counterions. For discrete charge
patterns of monovalent sites on the surface, the resulting
osmotic virial coefficient $B_2$ is found to be a strikingly
non-monotonic function of $c_s$.  The non-monotonicity follows from a
subtle Coulomb correlation effect which is completely missed by
conventional non-linear Poisson-Boltzmann theory and explains various
experimental findings.
\end{abstract}
]
\renewcommand{\thepage}{\hskip 8.9cm \arabic{page} \hfill Typeset
using REV\TeX }
\narrowtext
A more fundamental understanding of the interactions between
nano-sized biomolecules is critical to the long-term advance of modern
biomedical research \cite{Dill99}. The best strategy for a predictive
calculation is to study simple coarse-grained models where effects can
be clearly separated and approximations can be systematically
tested. While for micron-sized colloidal particles such coarse-grained
models have led to a quantitative understanding of the effective
interactions\cite{Isra92}, the challenging question is how far this
concept can be transferred to nano-particles.

A particular issue is the aggregation and crystallization of globular
proteins in solution, driven by their mutual interactions, including
steric repulsion, van der Waals attraction, Coulombic interactions,
hydration forces, hydrophobic attraction and depletion
forces\cite{Isra92}.  Most of these are effective interactions which
depend sensitively on solution conditions.  In particular Coulombic
forces are functions of pH (which determines the total charge of the
proteins) and of electrolyte concentration, which controls the Debye
screening length $\lambda_D$, and hence the effective range of
Coulombic interactions.  This dependence on solution conditions is
exploited in ``salting-out'' experiments where large salt
concentrations are used to trigger protein crystallization, a crucial
step towards the determination of their structure by X-ray
diffraction\cite{Dubr96}. While the forces acting between micro-sized
colloidal particles are dominated by generic interactions, and are
directly measurable by optical means\cite{Kepl94,Lars97,Verm98}, the
interactions between globular proteins are highly specific at short
range, and are less directly accessible.  One possible indirect
determination of the total force between two proteins may be achieved
via measurements of the osmotic equation of state by static light
scattering, which in the low protein concentration regime yields the
value of the second osmotic virial coefficient
$B_2$\cite{Geor94,Rose96}. 
The variation of $B_2$ with solution conditions yields
valuable information on the underlying effective protein pair
interactions.  Moreover, it has been shown empirically that there is a
strong correlation between the measured values of $B_2$ and the range
of solution conditions under which protein crystallization is
achieved\cite{Geor94,Rose96,Rose95,Vlie00}.

This report focuses on the effective interactions between globular
proteins mediated by the microscopic co and counterions, and on the
resulting $B_2$.  The conventional Derjaguin-Landau-Verwey-Overbeek
approach \cite{Verw48}, borrowed from colloid science, leads one to
expect that $B_2$ will monotonically decrease as the concentration of
salt increases, since higher salt concentrations lead to enhanced
screening (i.e. reduction of $\lambda_D$), and hence to a decrease of
the effective protein diameter.  This behavior rests on the standard
``coarse-grained'' model of uniformly charged colloids and smoothed
local densities of the microions.  We show that the discrete nature of
the protein surface charge distribution, together with the Coulomb
correlations between all charges involved, lead to a striking
non-monotonic variation of $B_2$ with salt concentration
$c_s$. 
 The occurrence of a minimum of $B_2$ as a
function of $c_s$ has recently been reported in lysozyme solutions for
$c_s =0.3$ M \cite{Guo} and in Apoferritin solutions for $c_s =0.15$ M
 \cite{Pets00}. Related experimental findings are
non-monotonic variations of other quantities which strongly correlate
with $B_2$ \cite{Guo,Bonnete} such as the interaction parameter
\cite{Grigsby,Mikol},the cloud point temperature \cite{Wu,Broi96}, and
the solubility \cite{Arakawa}. All these trends can be
qualitatively understood by our calculation.
\begin{figure}
\epsfxsize=6.cm 
\epsfysize=4.5cm
\vskip-0.25cm \epsfbox{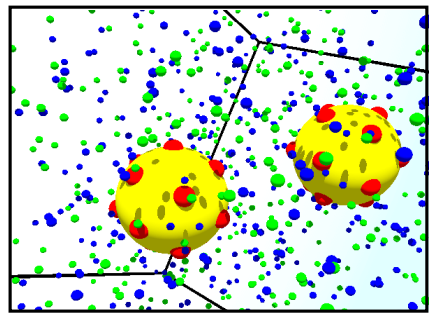}
\caption{\label{fig1}
Snapshot of a typical MD-generated microion configuration around two
proteins,
separated by $r=1.7 \sigma_p$.  The proteins carry 15 discrete charges
$e$; monovalent salt molarity is $c_s = 0.206 Mol/l$.  The globular protein
molecules are shown as two large gray spheres.  The embedded small
dark spheres on their surface mimic the discrete protein charges in
the DCM model. The small gray spheres are counterions, while the black
spheres are coions.}
\end{figure}
\vskip-0.25cm We consider two spherical proteins of diameter $\sigma_p$, each
carrying a total charge $Z e$ (where $Z$ depends on pH), surrounded by
monovalent co and counterions, assumed to have identical diameters
$\sigma_c$.  The solvent (water) is assumed to be a dielectric
continuum of permittivity $\epsilon$; this simplification, which
ignores the molecular granularity of the solvent, amounts to the
standard ``primitive'' model of ionic solutions\cite{Frie62}.

\begin{figure}
\epsfxsize=8cm 
\epsfysize=6.5cm
\vskip-0.25cm \epsfbox{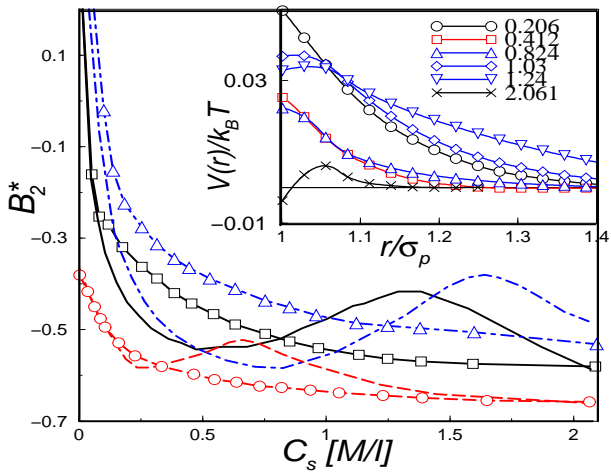}
\caption{\label{fig2}
 Normalized second virial coefficient $B_2^* = B_2/B_2^{HS}$ of a
protein solution versus added salt molarity.   Results are shown for
protein charges $Z=6$ (dashed lines), $Z=10$ (solid lines) and 
$Z=15$ (dot-dashed lines).  The lines with (without) symbols correspond to the
SCM (DCM) model.  The inset shows the effective
protein-protein interaction $V(r)$ in the DCM model versus separation
distance $r$ for $Z=10$.   Various symbols in the inset relate to the
different added salt molarities, indicated in the legend.} 
\end{figure}

In the case of highly charged colloidal particles, the total charge $Z
e $ is usually assumed to be uniformly distributed on the surface, a
situation which will be referred to as the ``smeared charge model''
(SCM).  This simplification is much less justified for the smaller,
weakly charged proteins (where $Z \simeq 10$).  We have hence
adopted a second, discrete charge model (DCM) where $Z$ monovalent
discrete point charges are distributed over the surface of a sphere of
diameter $\sigma_p - 2   \Delta = 0.96 \sigma_p$ (i.e.\ slightly inside the
surface of the protein), in such a way as to minimize the electrostatic
energy of the distribution; the resulting pattern does not correspond
to the real charge distribution on any specific protein, but does provide
a well-defined discrete model for comparison with the SCM, and between
different values of $Z$.  At this stage the two models (SCM and DCM)
involve only excluded volume and bare Coulombic interactions (reduced
by a factor $1/\epsilon$ to account for the solvent) between all
particles (proteins and microions).  However, in view of the large
size asymmetry, the effective force between the proteins, which
ultimately determines the second virial coefficient, involves a
statistical average over microion configurations in the field of two
fixed proteins\cite{Alla98}.  For distances $r > \sigma_p$ between the
centres of proteins $1$ and $2$, the total force $\vec{F}_1 =
-\vec{F}_2$ acting on each of the proteins is the sum of three
contributions, $\vec{F}_1 = \vec{F}_1^{(1)} + \vec{F}_1^{(2)} +
\vec{F}_1^{(3)}$, where $\vec{F}_1^{(1)}$ is the direct Coulomb
repulsion between the charges on the two proteins, $\vec{F}_1^{(2)}$
is the microion induced electrostatic force and $\vec{F}_1^{(3)}$ is
the depletion force due to the imbalance of the microion osmotic
pressure acting on opposite sides of the proteins\cite{Alla98}.  Both
$\vec{F}_1^{(2)}$ and $\vec{F}_1^{(3)}$ are averages over microion
configurations; according to the contact theorem $\vec{F}_1^{(3)}$ is
directly related to the integral of the microion contact density over
the surface of the protein\cite{Atta89,Pias95}.  The statistical
averages leading to $\vec{F}_1^{(2)}$ and $\vec{F}_1^{(3)}$ were
computed using Molecular Dynamics (MD) simulations.  The two proteins
were placed symmetrically with respect to the centre along the body
diagonal of a cubic simulation cell of length $L= 4 \sigma_p$, which
also contained monovalent co and counterions in numbers determined by
their bulk concentrations; periodic boundary conditions were adopted.
The choice of $L$ was made to ensure that the box length is much
larger than the range of the total (effective) protein-protein
interaction, so that the results would be independent of $L$.  For our
model to be a rough representation of lysozyme, we chose $\sigma_p = 4
nm$, and $Z = 6, 10$ and $15$, corresponding to three different
values of the solution pH. The microion diameter is $\sigma_c = 0.267
nm$.  Note that the SCM always implies vanishing multipole moments,
whereas within the present DCM, the only charge pattern with a
non-vanishing dipole moment is that for $Z=15$.  
A snapshot of a
typical equilibrium microion configuration around two proteins is
shown in Fig.\ref{fig1}, for the case $Z = 15$.  
Note that the dimensionless
Coulomb coupling parameter for a protein-counterion contact, namely
$\Gamma =  e^2 /[\epsilon k_B T( \Delta + \sigma_c/2)]$ for the DCM,
and $\Gamma = 2 Ze^2/[\epsilon k_B T(\sigma_p + \sigma_c)]$ for the
SCM, are comparable and of the order of $\Gamma=3$ at room
temperature.  The total force $\vec{F}_1 = -\vec{F}_2$ depends only on
the centre-to-centre distance $r$ for the SCM, but is also a function
of the orientations of the two charge patterns of the DCM, embodied by
two unit vectors $\vec{\omega}_1$ and $\vec{\omega}_2$; $\vec{F}_1 =
\vec{F}_1(r,\vec{\omega}_1,\vec{\omega}_2)$.  The anisotropy of the
force turns out to be relatively weak.  The effective radial pair
interactions between proteins, $V(r)$, follow from integration of the
radial projection of an orientationally averaged force $\vec{F}_1$
along the centre--to--centre vector $\vec{r}$, according to:
\begin{equation}\label{eq1}
V(r) = \int_r^\infty dr' \langle \frac{\vec{r}}{|\vec{r}|}\cdot \vec{F}(r',\vec{\omega}_1,\vec{\omega}_2)\rangle_{\omega_1
\omega_2}. 
\end{equation}
Here $\langle \ldots \rangle_{\omega_1,\omega_2}$ refers to a
statistical average over mutual orientations of the two
proteins\cite{rotate}.  The second virial coefficient in units of its
value $2 \pi \sigma_p^3/3$ for hard spheres of diameter $\sigma_p$,
$B_2^* = B_2/B_2^{(HS)}$, can then be proven to be given by:
\begin{equation}\label{eq2}
B_2^* = 1 + \frac{3}{\sigma_p^3} \int_{\sigma_p}^{\infty} dr r^2 
\left[1 - \exp \left\{ - V(r)/k_B T \right\} \right],
\end{equation}
a result formally identical to that valid for spherically symmetric
forces.  Results for $B_2^*$ as a function of salt concentration are
shown in Fig.\ref{fig2} for the SCM and DCM models, with three values of the
total protein charge.  In order to obtain values of $B_2$ comparable
to measured virial coefficients, we have taken short-range attractions
between proteins into account, by adding to the effective Coulomb
potential in Eq.~(\ref{eq2}) a ``sticky'' hard-sphere potential of the
Baxter form\cite{Baxt68}, with potential parameters $\delta = 0.02
\sigma_p$ and $\tau = 0.12$, which are known to yield reasonable
osmotic data for lysozyme solutions\cite{Rose95,Piaz98} in the high
salt concentration regime, where Coulombic interactions are
essentially screened out.

The key result, illustrated in Fig.\ref{fig2}, lies in the considerable {\em
qualitative} difference between the predictions of the SCM and the DCM
models for the variation of $B_2^*$ with monovalent salt concentration
$c_s$, irrespective of the total protein charge $Ze$.  While the SCM
(dashed curves) predicts a monotonic decay of $B_2^*$ with $c_s$, the
DCM leads to a markedly non-monotonic variation, involving an initial
decay towards a minimum followed by a subsequent increase to a maximum
and a final decrease towards a high $c_s$ value similar to that
predicted by the SCM.  The location of the minimum and of the maximum
shift to higher values of $c_s$ for larger protein charges $Z$.

The origin of the non-monotonic variation of $B_2^*$ with $c_s$ can be
traced back to the dependence of the effective (screened) Coulomb
interaction on salt concentration as shown in the inset of Fig.\ref{fig2}
for $Z=10$.  While the spherically averaged, repulsive effective
potential $V(r)$ of the DCM is initially strongly reduced as $c_s$ is
increased, its amplitude and range increase very significantly at
intermediate concentrations ($c_s \simeq 1M/l$), before it nearly
vanishes at the highest salt concentrations.  Note that $V(r)$ becomes
even slightly attractive at contact ($r=\sigma_p$) for $c_s \simeq
2M/l$.  The enhanced effective Coulomb repulsion at intermediate salt
concentrations cannot be rationalized in terms of simple mean-field
screening arguments; it is caused by a subtle correlation effect which
leads to the non-monotonic behavior of $B_2$ within the DCM.  The
protein-microion correlations are of a sufficiently different nature
in the SCM, to lead to a much more conventional, monotonic decay of
$B_2$ with $c_s$, similar to that expected from a linear screening
picture.
\vskip-0.25cm Even though the effective Coulomb potential between proteins is of
small amplitude, only a few percent of the thermal energy $k_B T$, the
effect on $B_2$ is dramatically enhanced by the presence of the strong
short-range attractive component due to van der Waals and hydrophobic
interactions, which we have included in the form of the Baxter
``sticky'' sphere potential.  This potential is independent of
salt-concentration, and has no influence on the qualitative dependence
of $B_2$ on $c_s$.

\begin{figure}
\epsfxsize=7cm 
\epsfysize=6.cm
\vskip-0.25cm \epsfbox{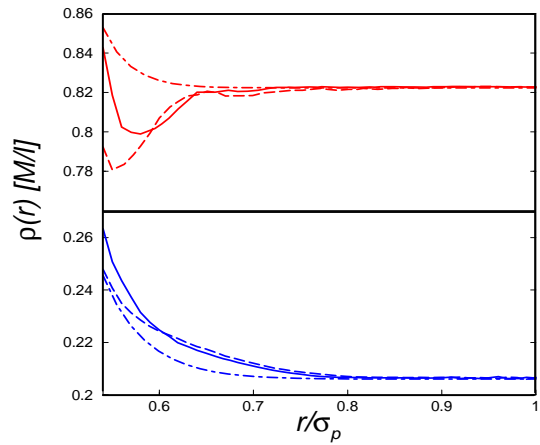}
\caption{\label{fig3}
 Total density profiles $\rho(r) = \rho_{+}(r) - \rho_{-}(r)$ of small
ions around a single protein, for salt molarities $c_s = 0.206$ (bottom
set of curves) and $c_s =0.824$ (upper set of curves).
 The solid and dashed lines
are simulation results for DCM and SCM models respectively, while the
dot-dashed lines are predictions of non-linear Poisson-Boltzmann theory.}
\end{figure}

\begin{figure}
\epsfxsize=7.cm 
\epsfysize=7.cm
\vskip-0.25cm \epsfbox{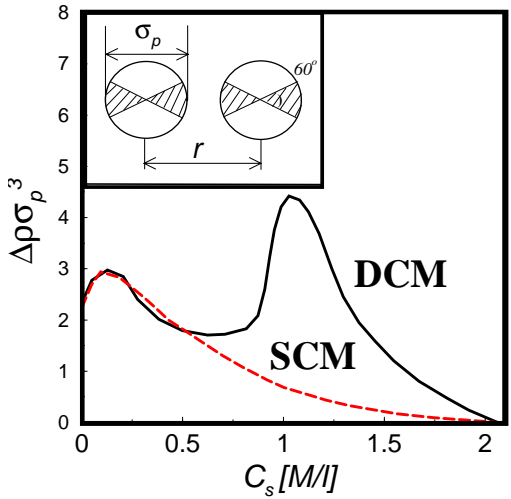}
\caption{\label{fig4} Microion density imbalance $\Delta \rho$ versus
salt molarity for protein charge $Z=10$ and separation $r=1.2
\sigma_p$.  The solid and dashed lines correspond to the DCM and SCM
models respectively.  The inset shows the angular range over which
$\Delta \rho$ is averaged (see text).} 
\end{figure}

In order to gain further insight into the physical mechanism
responsible for the unusual variation of the effective Coulomb
potential and of $B_2$ with salt concentration, we have investigated
in detail the local microion density in the immediate vicinity of the
protein surface.  The radial microion density profile $\rho(r) =
\rho_+(r) + \rho_-(r)$ around a single isolated protein is shown in
Fig.\ref{fig3}, for $Z=10$, and two salt concentrations (the profiles are
orientationally averaged in the case of the DCM).  At the lower salt
concentration ($c_s = 0.206 M/l$) the SCM and DCM models both yield an
accumulation of the microion density near contact, in
semi-quantitative agreement with the prediction of standard
Poisson-Boltzmann (PB) theory.  At the higher salt concentration,
however, there is a marked depletion in the microion density,
signaled by a minimum of $\rho(r)$ well below the asymptotic bulk
value.  This correlation effect is of course absent in the
(non-linear) PB theory, which always predicts a monotonically
decreasing density profile $\rho(r)$.  Note however a significant
difference between the SCM and DCM profiles.  While the latter
predicts a contact value $\rho_c(r=(\sigma_p + \sigma_c)/2)$ larger
than the bulk value, SCM predicts a much stronger microion depletion
near contact.  This finding illustrates the sensitivity of correlation
effects to the assumed charge pattern at the surface of a protein:
taking into account the discreteness of the surface charges leads to a
significant reduction of microion depletion at contact, compared to
the simplified picture of a uniformly smeared charge (SCM).
\vskip-0.25cm Next, consider the influence of a second near-by protein on the
microion distribution near contact.  We have computed the difference
between ``inner'' and ``outer'' shell microion contact densities, as
schematically illustrated in the inset to Fig.\ref{fig4}.  The local microion
density is no longer spherically symmetric, due to the interference of
the electric double-layers associated with the two proteins.  The
difference $\Delta \rho = \rho_{ in} - \rho_{ out}$
between the mean number of microions within a fraction of a spherical
shell of radius $R=0.6 \sigma_p$ subtended by opposite $60^{\circ}$ cones
is plotted in Fig.\ref{fig4} versus salt concentration.  $\Delta \rho$ is
always positive, indicating that microions (in fact mostly counterions)
tend to cluster in the region between the proteins, rather than on the
opposite sides, as one might expect due to the enhanced lowering of
the electrostatic energy for counterions shared between the two
proteins.  However, there is a very significant difference in the
variation of $\Delta \rho$ with $c_s$, between the SCM and the DCM
models.  Both exhibit similar behavior for $c_s \lesssim 0.5 M/l$,
with a small maximum around $0.2 M/l$.  Beyond $0.5 M/l$, however, the
SCM predicts a monotonic decrease of $\Delta \rho$, while the DCM
leads to a sharp peak in $\Delta \rho$ for $c_s \simeq 1 M/l$.  This
highly non-monotonic behavior clearly correlates with the
non-monotonicity of $B_2$ evident from Fig.\ref{fig2}. The excess number of
microions between the two proteins leads to an imbalance in osmotic
pressure, which is the origin of the increased {\em repulsion} between
proteins around $c_s = 1 M/l$, as shown in the inset of Fig.\ref{fig2}. For
stronger Coulomb coupling, as is the case for highly charged colloidal
particles in the absence of salt, the above depletion mechanism is
inverted, and leads to a depletion attraction between the
particles\cite{Alla98}, rather than to the enhanced repulsion found
here in the case of relatively weakly charged proteins.
\vskip-0.25cm The main finding of the present work is that the second osmotic virial
coefficient of protein solutions has a non-monotonic dependence on
salt concentration if the charge pattern on the protein surface is
discrete (as is the case for real proteins) rather than uniformly
smeared out, as usually assumed in the related case of
charge-stabilized colloidal dispersions, involving much larger
particles. The lesson to be learned from this finding is that one must
be cautious in attempting to extend coarse graining concepts and
approximations, developed and routinely used on the colloidal scale,
to the nanometric scale of proteins.  The discreteness of the charge
pattern is crucial to obtain non-monotonic behavior of $B_2$, which
is a subtle Coulomb correlation effect, totally missed by non-linear PB
theory.

We chose our simple models to help highlight and separate the effects
of discrete charge patterns and Coulomb correlations.  Extending our
MD calculations to the more complex (pH dependent) charge patterns of
realistic proteins\cite{Boye99} is technically straightforward.  We
expect that the physical mechanism leading to enhanced protein
repulsion at intermediate salt concentration, which is illustrated by
the microion density imbalance shown in Fig.\ref{fig4}, will carry over. 
Since the second osmotic virial coefficient determines much of the
excess (non-ideal) part of the chemical potential of semi-dilute
protein solutions, it is anticipated that the non-monotonicity of
$B_2$ may have a significant influence on protein crystallization from
such solutions in the course of a ``salting-out'' process.  The
non-monotonic behavior also suggests the possibility of an inverse,
``salting-in'' effect, whereby a reduction of salt concentration may
bring $B_2$ into the ``crystallization slot''\cite{Rose96}.

The authors are grateful to R. Piazza, I.L. Alberts,
P.G. Bolhuis, G. Bricogne, J. Clarke, S. Egelhaaf, J.F. Joanny, and
W.C.K. Poon for useful discussions, and to Schlumberger Cambridge
Research and the Isaac Newton Trust for financial support.
\end{document}